# Intelligent optoelectronic processor for orbital angular momentum spectrum measurement

**Hao Wang,[a,†] Ziyu Zhan,[a,†] Futai Hu,[a] Yuan Meng,[a] Zeqi Liu,[a] Xing Fu,[a,b,*] Qiang Liu[a,b,*]**
[a]Tsinghua University, State Key Laboratory of Precision Measurement of Technology and Instruments, Department of Precision Instrument, Beijing 100084, China
[b]Key Laboratory of Photonic Control Technology, Ministry of Education, Beijing 100084, China

**Abstract**. Orbital angular momentum (OAM) detection underpins almost all aspects of vortex beams' advances such as communication and quantum analogy. Conventional schemes are frustrated by low speed, complicated system, limited detection range. Here, we devise an intelligent processor composed of photonic and electronic neurons for OAM spectrum measurement in a fast, accurate and direct manner. Specifically, optical layers extract invisible topological charge information from incoming light and a shallow electronic layer predicts the exact spectrum. The integration of optical-computing promises us a compact single-shot system with high speed and energy efficiency (optical operations / electronic operations ~$10^3$), neither necessitating reference wave nor repetitive steps. Importantly, our processor is endowed with salient generalization ability and robustness against diverse structured light and adverse effects (mean squared error $\sim 10^{-5}$). We further raise a universal model interpretation paradigm to reveal the underlying physical mechanisms in the hybrid processor, as distinct from conventional 'black-box' networks. Such interpretation algorithm can improve the detection efficiency up to 25-fold. We also complete the theory of optoelectronic network enabling its efficient training. This work not only contributes to the explorations on OAM physics and applications, and also broadly inspires the advanced links between intelligent computing and physical effects.

[†]Equal contributing authors.
*Correspondence**:** Xing Fu, fuxing@tsinghua.edu.cn, or Qiang Liu, qiangliu@tsinghua.edu.cn

## 1 Introduction

Vortex beams carrying orbital angular momentum are ubiquitous in optical sciences. All OAM states of light constitute a Hilbert space, providing a new avenue to high-capacity communications[1], micro-manipulation[2], and quantum information processing[3]. Consequently, precise and efficient detection of OAM distribution or OAM spectrum becomes pivotal for these promising applications. Several previously proposed detection methods such as interferometry[4], diffractometry[5], and coordinate transformation[6, 7] are suitable for only detecting the existence of OAM within a limited range but inappropriate to extract accurate power distribution of OAM states. To this end, schemes based on mode projection and phase retrieval are explored[8-10], yet requiring not only many repetitive steps in data acquiring and postprocessing but also strict experimental calibration. Rotational Doppler effect is also employed to construct an OAM complex spectrum analyzer[11], at the cost of complicated detection setup and low speed. Another sequential weak and strong measurements in single-photon scenarios successfully reconstructs complex probability of 27-dimensional OAM states[12] but still fall short in terms of system conciseness and working speed. In short, despite of these realizations[4-9, 11-16], the limitations of detection speed, accuracy, range, robustness, generalization ability, and system conciseness hinder their practicability towards stable, fast and accurate information transfer in the modern age.

Along with enduring efforts in vortex beams comes the development of artificial intelligence. In particular, deep learning (DL)[17], has gradually revolutionized wide-ranging disciplines such as genetics[18], biomedical diagnosis[19] and physics[20]. In optics, data-driven DL algorithms are becoming pervasive tools to augment performance and to infuse new functionalities in imaging[21], holography[22], ultrafast photonics[23], optical metrology[24], etc. Recently, DL is also introduced to recognize OAM modes[25-28] due to their inherent ability to analyze complex patterns. However, the distinguishment of degenerate intensities patterns carrying different OAM spectra becomes a major challenge. More importantly, previous DL-related endeavors based on digital signal processing techniques often face poor generalization ability due to their 'black-box' nature and heavy computational consumption.

The flourishment of DL drives inconceivable demands for computing hardware especially at the era of big data. The computing power required to train or execute state-of-the-art DL models increases vastly while people's expectation on faster computing speed is still high. On the other hand, the development of integrated electronic circuits is unable to keep pace with the well-known Moore's law. Owing to the advantages of low latency, high energy efficiency and parallelism, photonics thus establishes itself in a central position when seeking alternative technologies for continued performance improvements[29, 30]. Many seminal photonic computing schemes are proposed recently[31-36]. Among them, wave-optics-based Diffractive Deep Neural Network ($D^2NN$) distinguishes itself with great flexibility[32, 37], depth advantage[38], and scalability[39]. In addition to many machine-vision demonstrations[32, 39] , it has been successfully incorporated into the diffuser imaging system[40], pulse shaping system[41], optical logic operation system[42], along with extension to different wavelengths[43-46].

Here, we demonstrate a single-shot measurement scheme called POAMS (processor for OAM spectrum) with high speed and interpretability, leveraging a hybrid optoelectronic neural network (Fig. 1). An optical diffractive network is synergized with a shallow electronic readout layer to predict the exact OAM spectrum for incoming structured light in a regressive manner. The obtained results on unknown experimentally generated single and multiplexed modes show that POAMS could be an optimal solution compared with the most advanced alternatives, featuring several critical properties: 1) high speed and energy efficiency: it works at microsecond level with most computation operations (~ 99.98% of all operations) optically conducted which cost little to no energy; 2) high accuracy: it can reconstruct sophisticated even random relative weights with mean squared error (MSE) around $10^{-5}$~$10^{-3}$; 3) conciseness: it entails neither reference wave nor repetitive measurements with a system size of ~$100\lambda \times 100\lambda \times 200\lambda$; 4) high robustness: it exhibits successful results even in the presence of adverse effects such as atmosphere turbulence and spatial dislocations; 5) great generalization ability: it functions well on experimental modes with diverse OAM spectra but without necessitating massive experimental training data; 6) great extendibility: it is implanted onto directly calculating OAM complex spectrum (i.e. the relative power and phase distribution). Moreover, we propose an efficient training method for optoelectronic models and demonstrate a universal model interpretation/visualization algorithm to comprehend the underlying physical mechanisms of our hybrid processor thus 1) removing the 'black-box' nature of neural networks and 2) benefitting system detection efficiency improvement. The synergy of optical and electronic neurons promises us a powerful and concise platform to facilitate OAM-based high speed information processing and to explore new opportunities and mechanisms of hybrid optoelectronic neural networks.

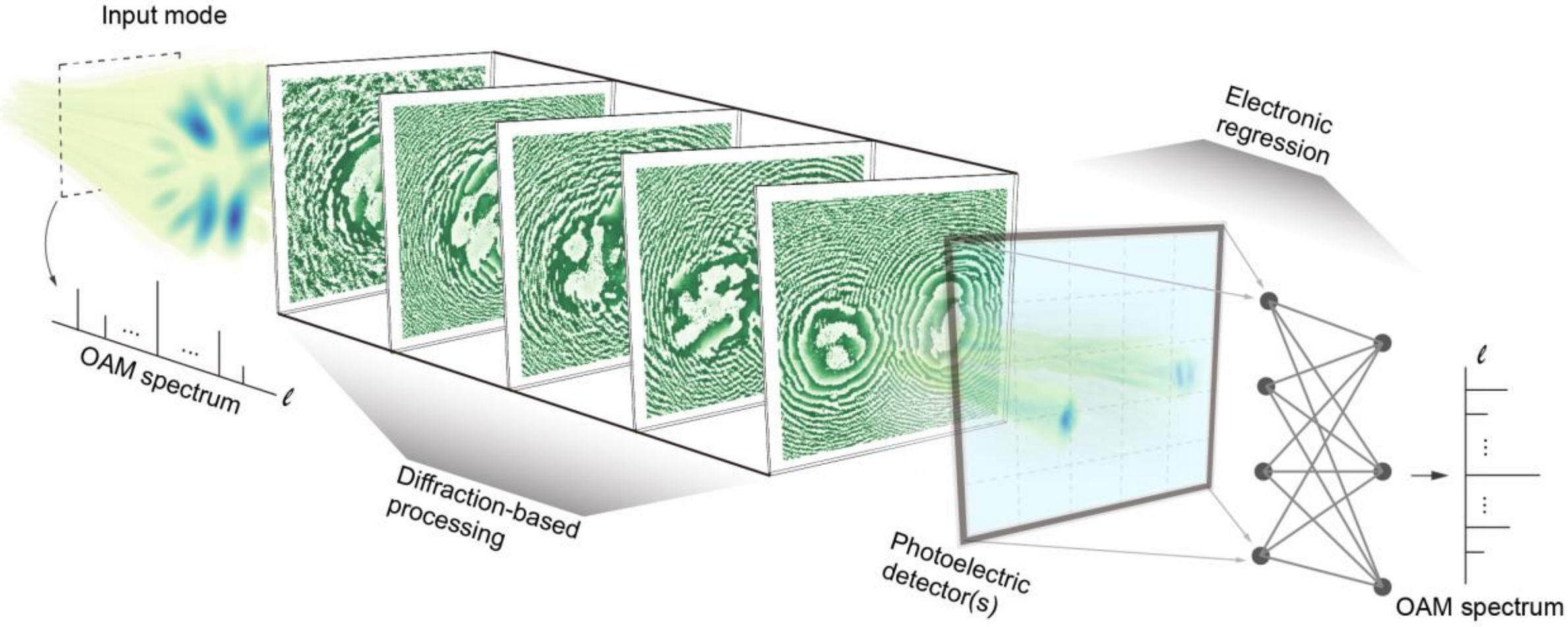

**Fig. 1** Illustration of the hybrid processor for OAM spectrum measurement. The incident structured light with certain OAM distribution is firstly processed by the diffractive optical neural network, whereby the OAM information is transformed into high-dimensional sparse feature in the photoelectric detector plane. For this model, the optical-diffraction-based processing part can 'split' the input beam into two main lobes that are related to positive and negative topological charges respectively. Once the complex optical field is turned into real-numbered intensity patterns, the shallow fully connected layer can recover the OAM spectrum in a regressive manner.

## 2 Results

### *2.1 Processor for OAM spectrum*

A general structured light $E(\rho,\theta,z)$ can be characterized by spatially mathematically decomposing onto orthogonal vortex modes:

$$E(\rho,\theta,z)=\sum_{\ell=K_N}^{K_P} c_\ell(\rho,z)\exp(i\ell\theta)\,, \tag{1}$$

where $(\rho,\theta,z)$ are cylindrical coordinates, $K_P$ and $K_N$ denote positive and negative OAM spectrum bounds for a given detection scenario (normally $|K_P|=|K_N|$ and they could be infinite), and $c_\ell(\rho,z)$represents complex coefficient function with respect to certain helical mode. Here we limit our attention to the degree of OAM by setting $c_\ell(\rho,z)=a_\ell \mathrm{LG}_{0,\ell}(\rho,z)$ with $\mathrm{LG}_{0,\ell}(\rho,z)=\mathrm{LG}_{0,\ell}(\rho,\theta,z)\exp(-i\ell\theta)$ representing the radial amplitude of Laguerre-Gaussian model $\mathrm{LG}_{0,\ell}(\rho,\theta,z)$ (radial index 0, azimuthal index $\ell$). Therefore, the normalized complex coefficient $a_\ell=|a_\ell|\exp(i\phi_l)$ contains OAM spectrum information where the amplitude of $a_\ell$ satisfies $\sum_{\ell=K_N}^{K_P}|a_\ell|^2=1$ and $\phi_l$ indicates intermodal phase with respect to a global reference phase. The OAM spectrum that elucidates the relative power distribution of all components can thus be expressed as a vector $\boldsymbol{s}=\left[\left|a_{K_N}\right|^2,\left|a_{K_{N+1}}\right|^2,\ldots,\left|a_{K_P}\right|^2\right]$. Calculating $\boldsymbol{s}$ from a given structured light is generally not easy and can be treated as an inverse problem. Figure 1 illustrates the workflow of our system to retrieve the OAM spectrum $\hat{\boldsymbol{s}}$. Accordingly, the whole picture of information processing in our scheme can be expressed as:

$$\hat{\boldsymbol{s}} = \mathcal{F}_E\left(\mathcal{F}_N\left(\mathcal{F}_O\big(E(\rho,\theta,z)\big)\right)\right), \tag{2}$$

with $\mathcal{F}_E$ and $\mathcal{F}_O$ the electronic regression function and the optical-diffraction-based wavefront transforming function respectively, and $\mathcal{F}_N$ the natural quadratic (nonlinear) function brought by photoelectric effect of a sensor. The optical neural network is composed of five cascaded phase layers with fixed distance ($40\lambda$) and each layer is endowed with $200^2$ programmable neurons. They serve as a special mapping to project the incident complex optical wave into a latent feature space once trained. The transformed complex features are switched to measurable real signals by the photoelectric sensor. Then the real features are fed into the shallow electronic fully connected layer (FCL) to obtain the spectrum. Note that earlier approaches involving all-optical $D^2NN$ or hybrid $D^2NN$ are normally applied for classification tasks in machine vision such as MNIST database[32, 47]. The POAMS in this work is distinctive from previous endeavors because OAM spectrum retrieval is a regression photonic problem in essence, which is rather challenging if only using optical neurons due to lack of nonlinearity. In this regard, the POAMS represents a new physical 'smart sensor'[35] in structured light processing in addition to recent demonstrations in fiber nonlinearity compensation[48] and optical computational imaging[40].

### *2.2 Results and analysis*

The proposed system architecture is shown in Fig. 1, where we craft the OAM spectrum analyzer during iteratively training with error backpropagation technique. The objective function can be expressed as:

$$\min_{\theta_o,\theta_E} \mathcal{L}(\boldsymbol{s},\hat{\boldsymbol{s}};\theta_o,\theta_E) + \mathcal{C}\|\theta_o\|_2^2 + \mathcal{C}\|\theta_E\|_2^2\,, \tag{3}$$

where $\mathcal{L}(\boldsymbol{s},\hat{\boldsymbol{s}};\theta_o,\theta_E)$ refers to the loss function comparing the processor's output $\hat{\boldsymbol{s}}$ and ground truth $\boldsymbol{s}$, and $\theta_o, \theta_E$ are optical and electronic neurons. The efficient training of hybrid networks is notoriously challenging[47], meaning that though parameters $\theta_o$ and $\theta_E$ are updated simultaneously, they are not balanced initially resulting in static optimization of $\theta_o$. We introduce two more hyperparameters for the smooth convergence of the hybrid model as shown in Fig. 2(a). The latter two terms with constant $\mathcal{C}$, known as $L_2$ regularization, are added as penalties to prevent overfitting and to increase model parameter sparsity.

The two adjacent loss curves in Fig. 2(a) validate that the model is not overfitting and the training ends after dozens of epochs. The optical processing part of our converged model is presented in Fig. 2(b). The resultant five diffractive layers mutually work to transform implicit OAM information into hierarchical features in the detector plane. Note that here these layers are not the same as those of Fig. 1 (trained with different hyperparameters and datasets), and the following test results are all based on the system in Fig. 2(b) instead of Fig. 1. Inspired by the beam steering effect induced by phase gradients (e.g. a focus lens), we analyze the fifth layer's gradient values in Fig. 2(b6) to straightforwardly show what exactly the diffractive layers are doing from the perspective of optics. It turns out that as the structured light propagates and interacts with these layers consecutively, the spatial wavevectors are mixed sufficiently in an 'intelligent' manner that the light fields are redirected towards different regions of the sensor plane. We repeat the training process several times under different hyperparameters and conclude that in most cases the incident optical field is transformed into speckle-like pattern, which contains high-dimensional features related to topological charges (TCs). Yet, an interesting case we obtain is the scheme sketched in Fig. 1: the optical part can 'split' the input beam into two main spatial lobes, where one lobe determines the OAM spectrum components with positive TCs and the other determines those with negative TCs. This phenomenon indicates that our optical neural network is actually transforming the wavefront of structured field. In this way, our scheme can naturally get rid of the misleading brought by degenerate intensity patterns, as confronted by earlier works using intensity-based DL recognition algorithms[25, 26, 28, 49].

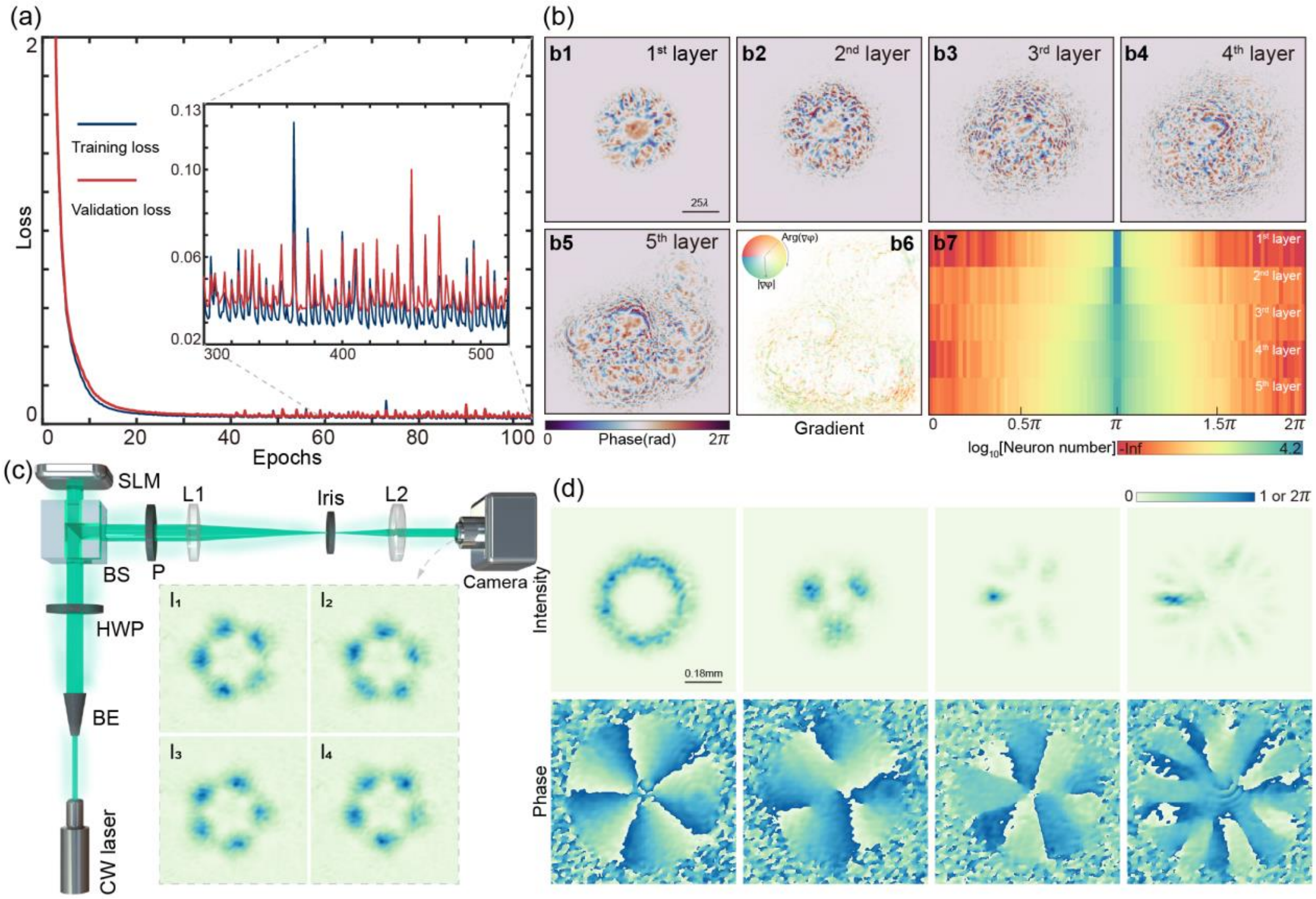

**Fig. 2** Training results and experimental data collection. (a) The loss curves of training set and validation set versus updating epochs. The learning rate is tuned dynamically every 15 epochs. The POAM converges after dozens of epochs and the inset indicates that the neural network is not overfitting. The loss value is the sum of 300 (one batch) samples. (b) Final designs of the optical diffractive network. (b1 - b5) Five cascaded diffractive layers with a fixed distance of 40λ between two successive layers. (b6) The gradient distributions of the 5th layer. PS: Color-encoded gradient map. (b7) Phase value distributions of 5 diffractive layers. (c) The optical setup for generating experimental structured light. CW laser, continuous-wave laser; BE, beam expander; HWP, half-wave plate; BS, beam splitter; SLM, spatial light modulator; P, polarizer; L1, L2, lens. For each data sample, we reconstruct the complex optical fields using 4-step phase shift method as depicted in the inset ($I_1$, $I_2$, $I_3$, $I_4$). (d) Four selective reconstruction results based on (c), which are single mode (TC = 5), multiplexed (mul.) mode (TC = $-4$, $-1$, equal weights), mul. mode (TC = $-4$, $-3$, 2, 5, equal weights), mul. mode (TC = $-10 \sim 10$, random weights), from left to right.

Furthermore, we calculate the statistical distributions of the optical (Fig. 2(b7)) and electronic neurons. All optical neurons are initialized with value $\pi$. After training, though most phase values are close to $\pi$, especially for the first layer, the gradual trend from the 1st layer to the 5th layer reflects that the interaction region between the layer and local optical field expands bigger. This quantitative observation not only indicates that the optical part is mixing the wavevectors towards the final target but also shows that the values of both optical and electronic neurons are sparse. In neural network studies, parameter sparsity benefits the feature selection and model interpretability[50]. In optics, this sparsity could manifest the convenience of physical fabrication/implementation of our engineered layers as well.

Next, we present numerical and experimental results from the POAMS for different tasks. Though the optical layers are implemented *in silico*, various structured light is experimentally generated as blind test samples to validate the reliability of trained model. We first reconstruct the complex optical fields through the 4-step phase shift method using the experimental setup in Fig. 2(c), which is also adopted in Ref.[8] for OAM spectrum detection. Some representative results are depicted in Fig. 2(d). We can clearly see the intensity and phase imperfections brought by nonideal laser source, possible astigmatism and distortion, sensor's noise, and other sources of error in our setup. Even so, the inference results from the processor are encouraging as we will exemplify in Fig. 3.

We first investigate the blind test performance on single (pure) vortex modes. We average the results from 30 repeated experiments and show them in Fig. 3(a). The dominant diagonal values indicate that the processor outputs nearly perfect OAM spectra with experimental single modes input, only with a little deviation in Gaussian modes. Then the processor is employed to measure the spectra of multiplexed OAM beams and two typical results based on 30 repeated experiments are illustrated in Fig. 3(b), where sample 1 denotes multiplexed beam with equal weights and sample 2 represents beam with random weights at the OAM bases. We can see a decent match between the output results and ground truths, even for very complicated OAM distributions. To further evaluate the measured results quantitatively, we calculate the R-squared ($R^2$) and mean squared error (MSE) values between the outputs and corresponding ground truths, as shown in Fig. 3(c). The average $R^2$ values for single modes, multiplexed modes with equal weights and random weights are 0.9924, 0.9268 and 0.8409, respectively, while the averaged MSE values are calculated as $2.878 \times 10^{-4}$, $4.822 \times 10^{-4}$, $1.617 \times 10^{-4}$ respectively, validating the success of POAMS to detect OAM spectrum. Interestingly, the $R^2$ metric reveals that the model performs better on single modes while the MSE metric shows better results on multiplexed modes with random OAM weights. This can be partly explained by the better generalization performance, due to the fact that the MSE metric and simulated multiplexed modes are employed as the loss function and the training dataset. Beyond that, we find the hyperparameter $T$ (also called temperature value) at the readout layer can sharpen/smoothen the OAM output. To further explore the influence of $T$, we train another 11 models with different $T$ values (from 0.01 to 1, logspace, each model trained with 60 epochs) and present the averaged MSE values upon different experimental datasets in Fig. 3(d). Indeed, $T$ can impact the convergence of the models as well as the final performance. With an optimized value e.g. $T \sim 10^{-1.2}$, one can achieve better results on multiplexed modes or single modes or mixture of them.

In addition, as the complex spectrum (weight $|a_l|^2$ and relative intermodal phase $\phi_l$) is reconstructed, the complex OAM state may be fully attained with prior knowledge of superposition principle for analyzing the in-depth optical properties (e.g. beam quality $M^2$, wavefront at any longitudinal distance)[9, 11, 51]. We extend our processor to regress the OAM complex spectrum by retraining a hybrid neural network with two parallel FCLs, which are responsible for the readouts of weights and relative phases respectively. Note the extension is rather implementable without losing the system's speed and energy advantages. The blind test results in Fig. 3(e) – (f) on simulated structured light are quite promising.

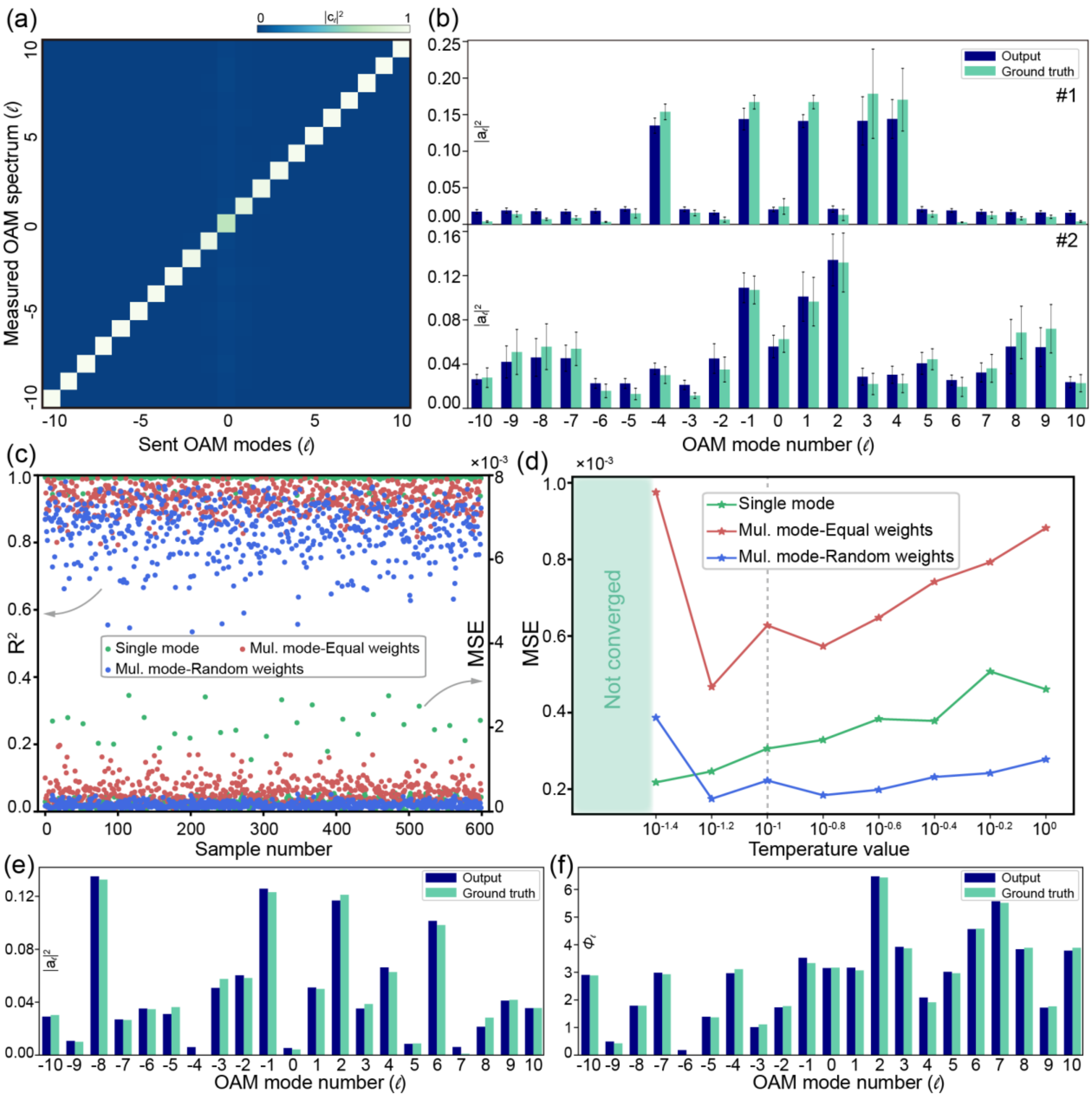

**Fig. 3** OAM spectrum results for different structured light. (a) The results of experimental single modes with TC from −10 to 10. Horizontal axis denotes different modes under test and longitudinal axis represents measured OAM spectrum. (b) Two selective results of experimental mul. modes. #1: mul. mode with equal weights. #2: mul. mode with random weights. Results of (a) and (b) are calculated based on 30 repeated experiments. Errorbar represents standard deviation. (c) Quantitative metrics to assess OAM spectrum measurement performance on different experimental datasets. Horizontal axis: index of each sample. $R^2$: R-squared. MSE: mean squared error. (d) Model performance comparison on different datasets versus temperature value $T$. Each MSE value is calculated through the average on the whole experimental dataset. Each $T$ correspondes to a trained model. The opaque cyan region means that the models with corresponding $T$ values are not converged during training. The vertical gray line marks the model in Fig. 2(a) – (b). (e) – (f) OAM complex spectrum on simulated mode. Left: weight. Right: relative phase.

### 2.3 Robustness against several adverse effects

In realistic circumstances, incident structured light is prone to be distorted by many uncontrollable adverse factors. Consequently, pronounced channel crosstalk (or OAM redistribution) arises, which is detrimental for practical communication links[1, 52]. Therefore, the robustness of an OAM spectrum detection system is of great significance[53]. Here, we quantitatively analyze how the POAMS reacts when facing spatial dislocations including transverse rotation (TR), longitudinal shift (LS), transverse shift (TS), angular shift (AS) and atmosphere turbulence (AT), as illustrated in the left panels of Fig. 4. The system is robust against these conditions if it can output the OAM spectra as the distorted ground truths. We evaluate our model on diverse test sets including pure vortex modes, multiplexed modes with equal intermodal weights or random intermodal weights.

**A. Transverse rotation.** It usually occurs when the detection module is indeliberately rotated. Additionally, as the optical layers in Fig. 1 and Fig. 2(b) possess no rotation symmetry, it's necessary to explore whether the POAMS is rotation-invariant. As shown in the right panel of Fig. 4(a), when the rotation angle varies from $-\pi$ to $\pi$, the averaged MSE errors barely change ($10^{-5}\sim10^{-4}$). Some randomly selected spectrum results are displayed in the inset, implying that the POAMS exhibits strong robustness against TR.

**B. Longitudinal shift.** Without changing OAM spectrum, distance change between the generation and detection module (i.e. longitudinal shift) will cause the deformation of incident wavefront induced by Gouy phase. Meanwhile, due to the divergence nature of propagating beams, LS also (de)magnifies beam width. As shown in Fig. 4(b), by changing LS from 0 to $1.0\ z_R$ (Rayleigh range), the output spectra maintain accurate, leading to distance-invariance of POAMS. It's mention-worthy that like any other optical systems, our model has an effective entrance pupil that can be observed from the interaction regions of the diffractive network as depicted in Fig. 2(b1). When the incident beam width exceeds the effective entrance pupil, there may be no sufficient modulation/processing and thus the processor loses efficacy.

**C. Transverse shift & Angular shift.** Another crucial factor is misalignment between generation and detection module due to transverse or angular shift[54], as shown in the left panels of Fig. 4(c) and (d). It is an essential hurdle because it is unavoidable in realistic setups. Unlike TR and LS which will not affect the ground truth OAM spectra, misalignment can induce distinct OAM spectrum changes[53]. Therefore, our preliminary results on single modes imply that there may exist mismatch between the outputs and ground truths. Specifically, the induced sidelobes of single mode spectra can be suppressed after the *softmax* output layer. To compensate for such mismatch, we leverage the flexibility of electronic neurons of the hybrid processor i.e. we conveniently retrain the electronic neurons (fine-tuning) with another 16000 distorted data samples (8000 for each) and keep the optical layers fixed. The blind test results are presented in Fig. 4(c) and (d). Owing to the TR robustness, we study TS varying from $-1$ to 1 (in beam width unit) and AS varying from $-9.1\times10^{-3}$ to $9.1\times10^{-3}$ (in rad) only in $xoz$ plane. The symmetric curves indicate identical performances.

**D. Atmosphere turbulence.** Besides the spatial imperfections, another critical factor that distorts the structured light during propagation is atmosphere turbulence (AT)[55]. To demonstrate the corresponding robustness, we put the POAMS into different AT environments. Specifically, the modified von Karman model is employed to generate random phase plates, where the outer and inner scales of turbulence are set as 10 m and 0.01 m respectively. AT refractive index structure parameter $C_n^2$ varies from $10^{-4.5}$ to $10^{-3}\ \mathrm{m}^{-2/3}$ and five different phase plates are used to accumulate 500 m propagation distance in turbulence. As TS and AS, AT also changes the OAM spectrum, therefore we also fine-tune the electronic weights. Test results from the retrained model is shown in the right panel of Fig. 4(e).

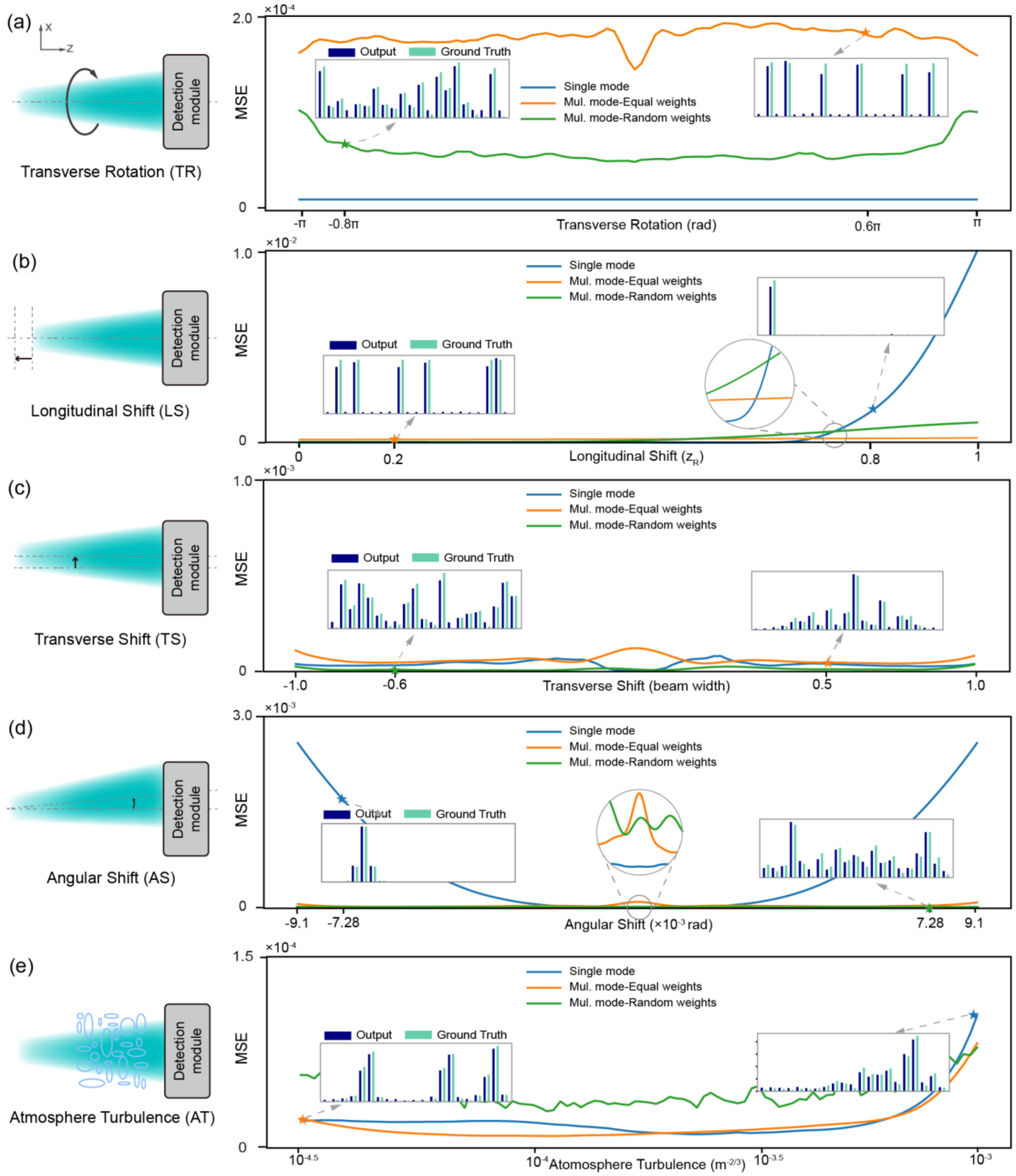

**Fig. 4** Robustness analysis of the POAMS against adverse conditions. Left panel: schematic diagrams of different adverse conditions. In the right panel, different curves represent the MSE values between POAMS outputs and ground truths after distortions on different datasets. Inset: randomly selected OAM spectrum results (TC range: $-10 \sim 10$) under corresponding adverse conditions. Every MSE value on curves is obtained from the average of 21 test modes. The test models in (c), (d) and (e) are fine-tuned with new electronic neurons to compensate for potential mismatches.

To sum up, the POAMS exhibits considerable robustness against above five adverse effects from the intuitional curves shown in Fig. 4. This robustness can further verify the wavefront transforming nature of the optical diffractive neural network. Namely our system is actually learning the global OAM information hidden in phase structures rather than memorizing the local trivial features. The analysis of LS manifests a large tolerance of effective entrance pupil of the system, which benefits high efficiency and signal-to-noise ratio. As for TS, AS and AT, the OAM spectrum changes after their distortions. Advanced fine-tuning techniques in neural networks can help us maintain the robustness of POAMS against them without compromising other advantages. Such robustness can effectively improve the practicability of OAM-based communication links and can inspire novel applications of vortex beams, e.g. the POAMS could act as a novel sensor for fast AT prediction[56]. Besides, we further investigate the robustness of the hybrid model itself including detector noise, interlayer distances and out-of-range mode inputs and put extended data.

### *2.4 Model interpretation*

As shown in Fig. 5(a), a ubiquitous convolutional neural network (CNN) is composed of several convolutional layers for extracting features and FCLs for combining learned features. Despite the popularity of CNN-enabled breakthroughs, a criticism is often raised on its 'black-box' nature, i.e. it is hard to explain why a given input produces a corresponding output[57]. On the other hand, clearly understanding how the neural networks work and what computations they perform can assist us to better improve the system. Otherwise, development of models is reduced to trial-and–error and misleading may arise without warning or explanation when models collapse[58-60]. To date, researchers have developed various techniques to visualize the CNN working principles e.g. t-SNE[61], CAM[62], and Grad-CAM[59]. Whereas, interpretation of hybrid optoelectronic neural networks still remains elusive. With the aid of optical diffraction theory[32] and previous visualization methods[58], here we propose a methos by: 1) analyzing the complex features behind each optical layer and 2) connecting the high-dimensional features in the sensor plane with the final OAM spectrum results.

For the POAMS, we treat diffractive layers as CNN's convolutional layers and also investigate the interaction between the last diffractive layer and readout layer. Specifically, we firstly present complex optical fields behind each diffraction-modulation unit to see how the incident structured light is processed optically. Figure 5(b) elucidates the hierarchical features that the optical neural network is extracting. Though the features seem rambling to us, they are critical to the network. Stated differently, the blended OAM information of structured light is transformed to easy-separable high-dimensional features in the latent space, so as to be globally regressed to model outputs. From our theoretical model, the latent space in the detector plane plays a significant role in communicating complex optical signals with real electronic signals. On
one hand, complex optical signals are turned to be real to provide greatly reduced parametric complexity. On the other hand, the compressed features enable us to interpret how each TC number connects to the spatial regions in the detector plane. Particularly, here we propose a method akin to occlusion sensitivity analysis in neural network studies[58], in accordance with the workflow in Fig. 5(c) and following steps:

i) Utilize a window ahead of the detector which only leaks a portion of signals to flow into the readout layer and occludes others.
ii) Monitor the consequence of OAM spectrum output and pick the OAM component which sticks

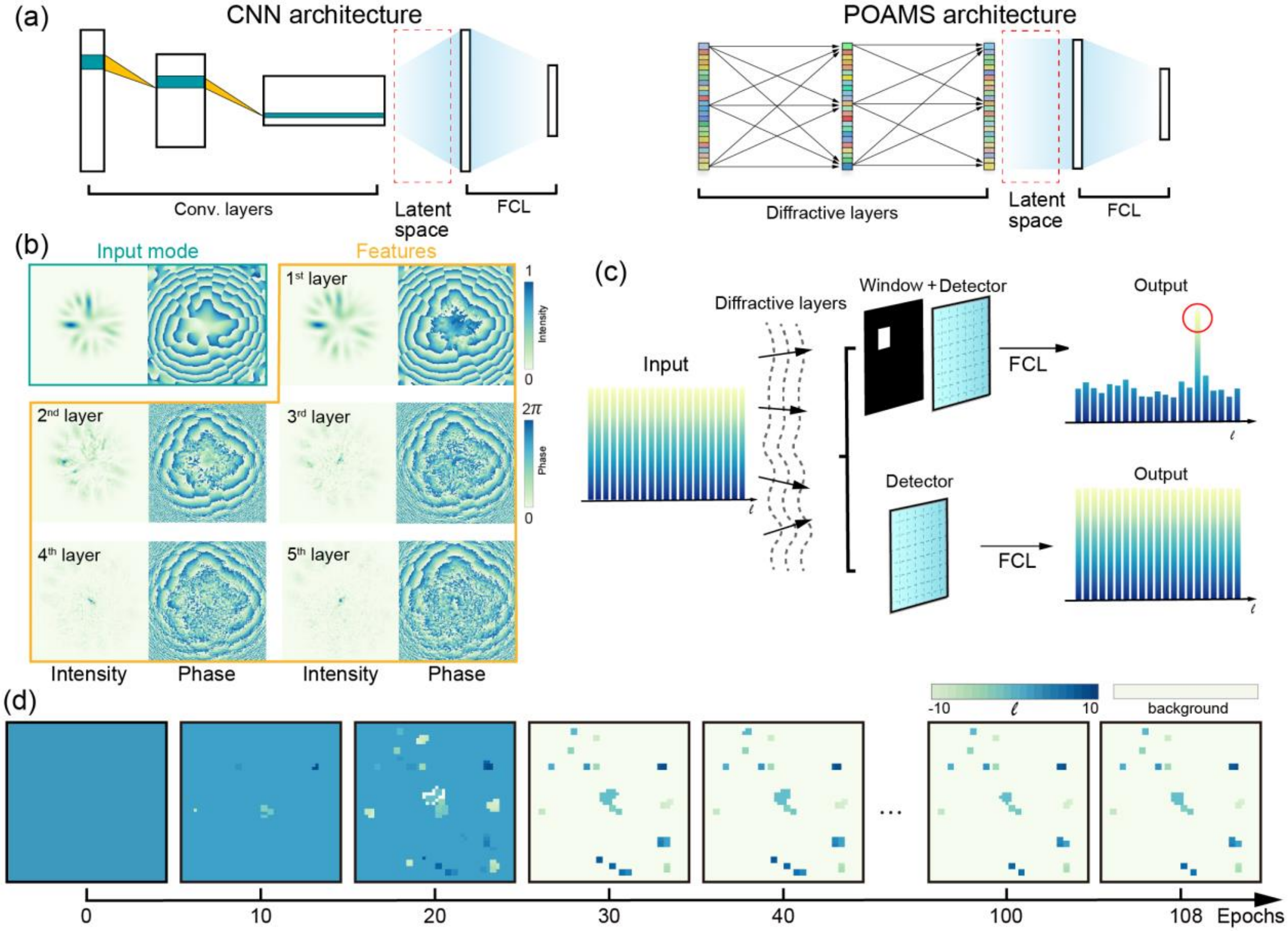

**Fig 5** Visualization of the POAMS. (a) Architecture comparison between a conventional CNN (left) and the POAMS (right) for further model interpretation. The red dashed rectangles represent the latent space connecting electronic convolutional layers (or optical diffractive layers) and FCLs. (b) Visualization of the complex optical features behind each diffractive layer. Intensity distributions are normalized. It can be seen that the diffractive layers can mix and redirect the spatial wavevectors 'intelligently'. (c) The visualization workflow for searching the specific spatial regions in the detector plane that connects corresponding TC number. Window size (white box size): $2 \times 2$, stride size: 1. (d) Evolution of obtained characteristic graphs during training. Colormap encoded with TC number. Model at #108 epoch is the best model (with the lowest validation loss).

out from others.

iii) Repeat step ii) for massive (e.g. 200) different input modes and decide one OAM component with maximum likelihood.

iv) Slide the window with a fixed stride across the whole plane and repeat steps i), ii) and iii) to obtain a feature map.

Here, we define such map that identifies the connection between feature spatial regions and exact OAM components as the *characteristic graph.* Without occlusion, the model results show high consistency with ground truth. With a sliding window, we obtain such graphs shown in Fig. 5(d). The characteristic graph exhibits several interesting properties. First, from the evolution of these graphs in Fig. 5(d), the distribution converges after adequate training (~ 45 epochs) i.e. the general structure remains stable, only with tiny changes along evolution. Second, the distribution of the graph is globally invariant against the window size. Third, the graph tends to be complete during training, meaning that the graph contains all 21 OAM components. More importantly, we find that once obtaining the graph, we are able to: 1) determine the existence of certain OAM component and 2) reconstruct the OAM spectra quite accurately, by only detecting the corresponding spatial regions at the graph. In other words, according to the graph, one can obtain the OAM information with much less scanning region and measurement steps when using a single pixel detector (photodiode) or with much less detector size when using an array, which significantly improves detection efficiency by 25-fold. This liberates us from 2D full screen demodulating devices[10] and detectors[8, 9] when measuring OAM to push the speed to the limit. Such phenomenon also reflects that the optically extracted features are sparse in the detector plane.

In short, visualization of this model helps us better understand the interaction mechanisms between layers. Especially, we can determine which part of intensity signals in the detector plane contributes to certain OAM component. In this regard, the POAMS is significantly suitable for structured light detection by optically transforming corresponding wavefronts, extracting decisive features, decoupling the mixed TC information and electronically combining the global features to the final results. Compared to all-optical or all-electronic neural networks, the POAMS reported in this work strikes a remarkable balance among model expressivity, application scenario, inference speed and energy efficiency.

## 3 Discussions

### *3.1 Model analysis*

The above results clearly show the exceptional power of the hybrid optoelectronic processor for OAM spectrum detection. First, the POAMS contains 0.2 million optical neurons and 5021 electronic neurons, leading to remarkable speed and energy efficiency. In other words, the optical-achieved $7.68 \times 10^{10}$ computation operations cost little to no energy and are performed nearly at the speed of light, and electronic-achieved $1.26 \times 10^{7}$ operations cost relative low energy using a moderate computer. Then, compared with existing OAM spectrum measurement methods, the POAMS functions in a direct single-shot manner, preventing the need of strict alignment for mode projection, reference wave for phase retrieval, etc. This, in one way, promises us a compact and elegant system (approximately $100\lambda \times 100\lambda \times 200\lambda$ with $\lambda$ the working wavelength), and in another way, saves time and energy.

For the neural network convergence, despite fruitful discussions in all-optical $D^2NN$ training, the instructional analysis of hybrid optoelectronic neural networks is elusive[47], especially in terms of smooth and direct training pipeline, which hampers the advanced applications of these models. In this work, we find the imbalance between optical neurons and electronic neurons. Specifically, the gradient-descent-based algorithm updates the electronic neurons effectively but the optical parameters stay static, thus hindering the successful updating of the joint model. By calculating the back propagation errors analytically, we introduce two hyperparameters to augment the errors with respect to optical neurons, yielding a synchronized network parameter update.

### *3.2 Implementation*

The overall system is composed of three parts: structured light generation, optical diffractive layers and electronic readout layer. For the first part, we experimentally generate various optical states as blind test sets to evaluate the POAMS. We emphasize that the generalization ability of POAMS is highly satisfactory. Typically, the applications of DL in optics are often hindered by the need of massive labelled experimental data to drive the training of the network parameters because the collection of substantial experimental data in optics is prohibitively time-consuming and sometimes impractical[63]. However, in this work even the model is trained *in silico* using limited simulated dataset (only multiplexed modes with random OAM weights), the obtained processor is capable of predicting experimental single modes and multiplexed modes with equal as well as random weights collected via the setup in Fig. 2(c) (see results in Fig. 3). Free from experimental training sets and great generalization ability are desirable for real-world applications by fast decoding various unknown OAM beams. For the second part, in addition to many insightful simulation studies on $D^2NN$[37, 38, 64-66], some experimental advances also demonstrate the feasibility of transferring *in silico* diffractive layers to real devices[32, 44-46], indicating that the computer-optimized models are also accurate and convincing. Note the working wavelength is decisive when physically implementing these layers. 3D printed layers have been unlocked in the terahertz spectral range[32], radio frequency range[45], and acoustic wave scheme[44], which are relatively large in size thus easy to fabricate. Considering vortex beams are mostly investigated in optical frequencies e.g. 532 nm, the fabrication of optical neurons is harder. Even so, if with sufficient laboratory sources, we can resort to multi-step photolithography-etching technique[46] or five cascaded SLMs[67] or metasurface[68] to optically extract TC features. To further improve the reliability, in this work we conduct the free space propagation simulation based on angular spectrum method with zero padding[69], which has been proven to be accurate enough in holography[70]. For the third part, since the electronic FCL is shallow and only covers limited computational operations, we use a moderate computer that is qualified to achieve high speed processing. More importantly, the FCL can mitigate the possible discrepancy between the experimental results and simulation counterparts through adaptive training technique[39]. Specifically, given 5 fabricated optical layers, one can compensate the imperfections arisen from fabrication or experimental setup e.g. misalignment by only fine-tuning the electronic FCL without changing any devices as demonstrated in above robustness section. This can be another advantage of our hybrid system. The codes and data are available for the whole implementation to train/test the system and for developing new processors at https://github.com/hao-focus/ModelForOAMLight.

### *3.3 Future improvement*

The presented POAMS is scalable and can be further improved. For example, precisely detecting structured light in the OAM basis in other optical spectra can be achieved utilizing this platform. Also, one can easily extend the OAM detection range by adjusting the number of neurons in the last layer, without compromising other advantages. To advance the results reported in Fig. 3 and Fig. 4, one can scale up the training dataset by adding more diverse data samples. Note we do not have to start from scratch to train the network, transfer learning technique[71] can be employed to fine-tune the POAMS. In this work, we limit our attention to scalar structured beams and the engineered diffractive layers are polarization-insensitive. Vectorial counterpart is also feasible[66, 68], which could inspire efficient vector structured light detection. In addition to azimuthal mode spectrum this work discussed, crafting the POAMS into a radial mode spectrum can be relevant future work[72]. Besides, the POAMS can be recognized as a solution to a regression problem in essence, and it could be extended to solve similar problems in other fields which necessitates high speed, accuracy and robustness.

## 4 Conclusions

In summary, we have demonstrated a compact optoelectronic processor for structured mode analysis, which can directly detect OAM spectrum of structured light in a fast, accurate and robust manner. Our processor allows one to immediately obtain the TC information without any interference measurements nor repetitive steps, empowered by the hybrid computing nature. We sharply ease the workload of deep-learning models on collecting massive experimental training data in obtaining a hybrid model with great generalization ability. We validate the performance on experimental and simulated modes with diverse (complex) OAM spectra even against nonideal conditions such as atmosphere turbulence, misalignment. In addition, we observe interesting connections between TC numbers and the optical neurons and consequently propose a universal model interpretation paradigm for hybrid neural networks, which not only help us understand the overall system but also further improve the detection efficiency. This study highlights the advantages of fusing optics and electronics in settling photonics problems through physical smart sensors. More specifically, this work closes a practical gap in OAM-based high-speed information processing and facilitates both OAM and optoelectronic neural networks studies.